\documentclass[aps,preprint,nofootinbib,superscriptaddress,prc]{revtex4}
\usepackage{amssymb}

\usepackage{epsfig}
\usepackage[bookmarksnumbered,bookmarksopen,colorlinks,citecolor=blue,linkcolor=blue] {hyperref}
\begin{document}

\date{\today}
Submitted to Chinese Physics C
\title{Effect of Wigner energy on the symmetry energy coefficient in nuclei}

\author{Junlong Tian}
 \email{tjl@aynu.edu.cn}
 \affiliation{School of Physics and Electrical Engineering,
Anyang Normal University, Anyang 455000, People's Republic of
China }

\author{Haitao Cui}
\affiliation{School of Physics and Electrical Engineering,
Anyang Normal University, Anyang 455000, People's Republic of
China }

\author{Teng Gao}
\affiliation{School of Physics and Electrical Engineering,
Anyang Normal University, Anyang 455000, People's Republic of
China }

\author{Ning Wang}\email{wangning@gxnu.edu.cn}
\affiliation{ Department of Physics, Guangxi Normal University,
Guilin 541004, People's Republic of China }
\affiliation{ State Key
Laboratory of Theoretical Physics, Institute of Theoretical Physics,
Chinese Academy of Sciences, Beijing 100190, People's Republic of
China}

\begin{abstract}
The nuclear symmetry energy coefficient (including the coefficient $a_{\rm sym}^{(4)}$ of $I^{4}$ term) of finite nuclei is extracted by using the differences of available experimental binding energies of isobaric
nuclei. It is found that the extracted symmetry energy coefficient $a^{*}_{\rm sym}(A,I)$ decreases with
increasing of isospin asymmetry $I$, which is mainly caused by
Wigner correction, since $e^{*}_{\rm sym}$ is the summation of
the traditional symmetry energy $e_{\rm sym}$ and the Wigner energy
$e_{\rm W}$. We obtain the optimal values $J=30.25\pm0.10$ MeV, $a_{\rm ss}=56.18\pm1.25$
MeV, $a_{\rm sym}^{(4)}=8.33\pm1.21$ MeV and the Wigner parameter $x=2.38\pm0.12$
through the polynomial fit to 2240 measured binding energies for
nuclei with $20 \leq A \leq 261$ with an rms deviation of 23.42 keV.
We also find that the volume symmetry coefficient $J\simeq 30$ MeV
is insensitive to the value $x$, whereas the surface symmetry
coefficient $a_{\rm ss}$ and the coefficient $a_{\rm sym}^{(4)}$ are very
sensitive to the value of $x$ in the range $1\leq x\leq 4$. The contribution of $a_{\rm sym}^{(4)}$ term increases
rapidly with increasing of isospin asymmetry $I$. For very
neutron-rich nuclei, the contribution of $a_{\rm
sym}^{(4)}$ term will play an important role.
\end{abstract}

\maketitle

\section{\label{sec:level2}Introduction}
It is evident that the symmetry energy coefficient plays an extremely
important role, not only in nuclear physics,  such as the dynamics
of heavy-ion collisions induced by radioactive beams, the proper description of the nuclear binding energies along the periodic table, and
the structure of exotic nuclei near the nuclear drip lines
\cite{Dani02,Stein05,Bara05,Latt07,liba08,Dong11,Rswang14,Ouli15}, but also a number
of important issues in astrophysics, such as the dynamical evolution
of the core collapse of a massive star and the associated explosive
nucleosynthesis
\cite{Latt00,Horo01,Todd05,Shar09,Kumar11,tianwd11,Fatto13}. In the
global fitting of the nuclear masses in the framework of the
liquid-drop mass formula, the symmetry energy per particle is
usually written as $e_{\rm sym}=a_{\rm sym}I^{2}$, in which the symmetry
energy coefficient $a_{\rm sym}$ enters as a mass-dependent
phenomenological parameter \cite{Janec03,Ono04,WangN2,Oyam10,Nik11}.
In fact, $a_{\rm sym}$ could also be a function of the isospin
asymmetry $I=(N-Z)/A$. The isospin dependence of the symmetry
coefficient $a_{\rm sym}$ is usually written as $a_{\rm
sym}(A,I)=J-a_{\rm ss}/A^{1/3}+a_{\rm sym}^{(4)}I^{2}$ by neglecting the
higher order term, the same as in Ref. \cite{Wang15,Jiang15}.
But how to change it depends on the isospin asymmetry $I$ for given mass number $A$, decreases or increases? It is mainly determined by the high-order $I^{4}$ term coefficient $a_{\rm sym}^{(4)}$ of the symmetry energy.
However the coefficient $a_{\rm sym}^{(4)}$ is
difficult to be determined. It is necessary to investigate
the symmetry energy coefficient of finite nuclei.

In Ref. \cite{Liu10}, Min Liu et al. obtained the mass dependence of $a_{\rm sym}(A)$
through performing a two-parameter parabola fitting to the energy
per particles after removing the Coulomb energy $e_{n}(A,
I)=e(A,I)-e_{c}(A,I)$ for a series of nuclei with the same mass
number $A$. The extracted $a_{\rm sym}$ is only dependent on mass
number $A$. In this work, with the similar approach in Ref. \cite{Liu10}, we consider the mass
and isospin dependence of $a_{\rm sym}$, and at the same time the
higher-order ($I^4$) term of the symmetry energy is included. It is
found that the Wigner energy $E_{W}$ should be considered in the
extraction of nuclear symmetry energy coefficient. The Wigner energy
can be extracted from the difference of $e_{n}(A,I)$ of
isobaric nuclei. However the Wigner energy is not included in
extracting the symmetry energy coefficient in our previous
paper \cite{Tian14}. The nature of the symmetry and Wigner energy are
intertwined in the nuclear mass formula and that one term cannot be
reliably determined without knowledge of the other \cite{Isack06}.
This leads to considerable uncertainty in the value for the symmetry
energy, especially the coefficient $a_{\rm sym}^{(4)}$ of the $I^4$
term in the symmetry energy coefficient expression.

The paper is organized as follows. In Sec. II,  the symmetry energy
and Wigner energy are described and the summation of both are
extracted by using the experimental binding energies differences
between isobaric nuclei. In Sec. III, The method of extracting the
symmetry energy coefficient is described and the corresponding
coefficients are obtained through the polynomial fitting. The effect of
the Wigner energy term on the symmetry energy coefficient is also
studied. in Sec. IV. Finally a summary is given in Sec. V.

\section{\label{sec:level1}Symmetry energy and Wigner energy}
It is well known that nuclear mass is one of the most precise
measured quantity in nuclear physics. It can provide information of
the symmetry energy coefficient through the
liquid-drop mass systematics. In semi-empirical
Bethe-Weizs$\ddot{a}$cker mass formula \cite{Weiz35,Bethe36}, the
energy per particle $e(A,I)$ of a nucleus can be expressed as a
function of mass number A and isospin asymmetry $I =(N-Z)/A$,
\begin{eqnarray}
e(A,I)=a_{v}+a_{s}A^{-1/3}+e_{c}(A,I)+a_{\rm sym}I^{2}+\delta,
\end{eqnarray}
with
\begin{eqnarray}
\delta=\pm a_{p}A^{-3/2} ~ or ~ 0,
\end{eqnarray}
where the ``+" is for even-even nuclides, the ``--" is for odd-odd
nuclides, and for odd-A nuclides (i.e. even-odd and odd-even)
$\delta=0$. The a$_{v}$, a$_{s}$, a$_{\rm sym}$ and a$_{p}$ are the
volume, surface, symmetry and pairing energy coefficients,
respectively. The Coulomb energy per particle is $e_{c}(A,I)=E_{c}/A$,
where the Coulomb energy of a nucleus
$E_{c}=0.71\frac{Z^2}{A^{1/3}}(1-0.76Z^{1/3})$ and
$Z=\frac{A}{2}(1-I)$ are usually used \cite{Wang10,Wang11}.

Let us assume the binding energy per particle
$e(A,I)=e_{n}(A,I)+e_{c}(A,I)$, $e_{n}(A,I)$ and $e_{c}(A,I)$ denote
the nuclear energy part and the Coulomb energy part per particle,
respectively. Subtracting the Coulomb energy term from the binding
energy, one obtains the nuclear energy part per
particle,
\begin{eqnarray}
\label{eq3}
e_{n}(A,I)&=&e(A,I)-e_{c}(A,I)\nonumber\\[1mm]
&=&e_{0}(A)+e_{\rm sym}(A,I)\nonumber\\
&=&e_{0}(A)+a_{\rm sym}(A,I)I^{2},
\end{eqnarray}
where $e_{0}(A)=a_{v}+a_{s}A^{-1/3}+\delta$ including the volume, surface
and pairing energy terms, is only dependent on nuclear mass
number $A$. $e_{\rm sym}(A,I)$ is the symmetry energy per
particle of a nucleus. If we take the difference in the nuclear
energy part per particle $e_{n}(A,I)$ between two isobaric nuclei
with same odd-even parity, $e_{0}(A)$ term is canceled and the
difference of the symmetry energy per particle can be written as
\begin{eqnarray}
\Delta e_{\rm sym}=e_{n}(A,I)-e_{n}(A,I_{1})=a_{\rm
sym}(A,I)I^{2}-a_{\rm sym}(A,I_{1})I_{1}^{2},
\end{eqnarray}
Here $e_{n}(A,I_{1})$ is the nuclear energy part per particle of a reference nucleus ($A,I_{1}$), and the symmetric nuclei ($I_{1}=0$) is selected as the reference nucleus if its experimental binding energy is exist for even-even nuclei. For any other case, the nuclei with the minimum value of $I_{1}=I_{min}>0$ is selected as the reference nucleus among each series isobaric nuclei. $e_{n}(A,I)$ is the any other value of isobaric nuclei for given mass number $A$.

If the experimental binding energy of a symmetric nucleus ($I_{1}$=0) is known,  we obtain
\begin{eqnarray}
e_{\rm sym}(A,I)=e_{n}(A,I)-e_{n}(A,0)=a_{\rm
sym}(A,I)I^{2},
\end{eqnarray}
or
\begin{eqnarray}
a_{\rm sym}(A,I)=\frac{e_{\rm
sym}(A,I)}{I^{2}}=\frac{e_{n}(A,I)-e_{n}(A,0)}{I^{2}},
\end{eqnarray}
where only even-even nuclei are taken into account in our calculations to consider the paring effects for the even mass number nuclei.
\begin{figure}
\includegraphics[angle=-0,width= 0.65\textwidth]{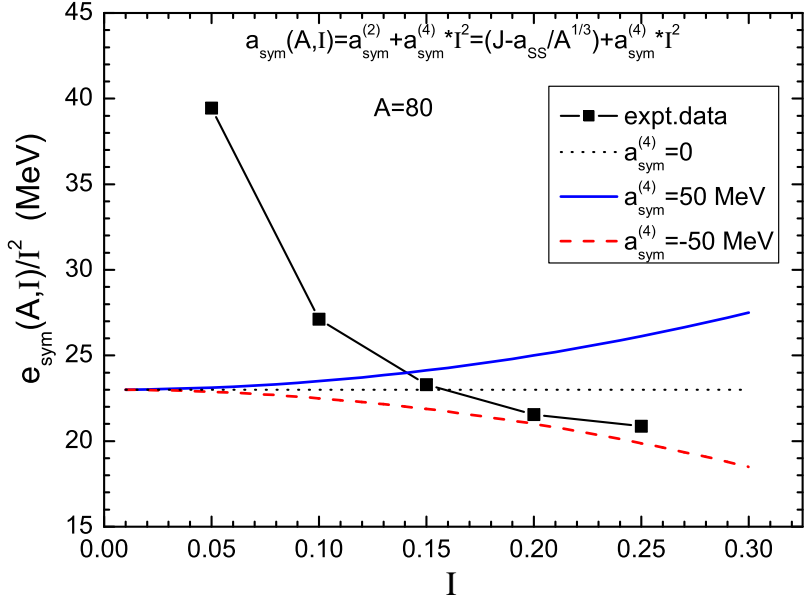}
 \caption{(Color online)Experimental symmetry energy coefficients as a function of $I$ extracted from Eq. (6) for all even-even nuclei with mass number $A$=80 (solid squares). The dotted line ($a_{\rm sym}^{(4)}$=0), the solid line ($a_{\rm sym}^{(4)}$=50 MeV) and the dashed line ($a_{\rm sym}^{(4)}$=--50 MeV) are the results using the expression of symmetry energy coefficient of Eq. (7).}
\end{figure}

On the other hand, according to the liquid drop model, the symmetry energy coefficient of a finite nucleus is usually written as
\begin{eqnarray}
\label{eq7}
a_{\rm sym}(A,I)&=&a_{\rm sym}^{(2)}+a_{\rm sym}^{(4)}I^{2}+o(I^{4})\nonumber\\[1mm]
&=&\simeq J-a_{\rm ss}A^{-1/3}+a_{\rm sym}^{(4)}I^{2},
\end{eqnarray}
by using the Leptodermous expansion in terms of powers of
$A^{-1/3}$. $J\approx28-34$ MeV denotes the symmetry energy of
nuclear matter at normal density. $a_{\rm ss}$ is the coefficient of the
surface symmetry term. $a_{\rm sym}^{(4)}$ is the coefficient of the
$I^4$ term in the expression of symmetry energy.

Figure 1 shows the extracted experimental symmetry energy
coefficients as a function of isospin asymmetry $I$ extracted from
Eq. (6) for all even-even nuclei with mass number $A$=80 (solid
squares), where $e_{n}(A,I)=e(A,I)-e_{c}(A,I)$, the experimental
binding energy per particle $e(A,I)$ is taken from the mass table
AME2012 \cite{AME2012}, and
$e_{c}(A,I)=0.71\frac{Z^2}{A^{4/3}}(1-0.76Z^{1/3})$. The dotted line
($a_{\rm sym}^{(4)}$=0), the solid line ($a_{\rm sym}^{(4)}$=50 MeV)
and the dashed line ($a_{\rm sym}^{(4)}=-50$ MeV) are the results
using the expression of symmetry energy coefficient Eq. (7) with
$a_{\rm sym}^{(2)}$=23 MeV. From figure 1, one can see that only
using the expression of symmetry energy coefficient of Eq. (7), the
extracted experimental symmetry-energy coefficient  can not be
reproduced whatever it is positive, zero or negative value for
$a_{\rm sym}^{(4)}$.

The effect of the Wigner energy is  responsible for the
decrease of $e_{\rm sym}(A,I)/I^{2}$ with isospin asymmetry $I$ at a
given mass number $A$. To reproduce the experimental data better,
one should include the Wigner energy term in Eq. (5). Let us rewrite the expression of
Eq. (5) as $e_{\rm sym}^{*}(A,I)=e_{n}(A,I)-e_{n}(A,0)$, where
$e_{\rm sym}^{*}(A,I)$ is defined as the summation of the
traditional symmetry energy $e_{\rm sym}(A,I)$ and the Wigner energy
$e_{W}(A,I)$. However the different Wigner energy expression and
parameters will directly affect the extraction of symmetry energy
coefficients. Figure 2 (a) presents two forms for Wigner energy which is a function
of isospin asymmetry $I$ and applied to all even-even nuclei with
mass number $A$=80 in mass table AME2012. One is
$e_{W}=29.156I^{2}[(2-|I|)/(2+|I|A)]$ (solid triangles), which is
proposed in Ref. \cite{Wang10}, the other is
$e_{W}=-10\exp(-4.2|I|)/A$ \cite{Myer97}(solid circles), which is
usually used in the literature. For convenience we denote the former
by ``form (1)" and the latter by ``form (2)", respectively. From
Fig. 2 (a) one can see that the value of $e_{W}$ is positive for form (1) and
negative for form (2). While the value $e_{\rm sym}^{*}(A,I)$ is the
summation of the traditional symmetry energy and the Wigner energy,
negative Wigner energy of form (2) will lead to a larger traditional
symmetry energy and thus larger symmetry energy coefficient than
that with form (1). Fig. 2 (b) presents the extracted symmetry-energy
coefficients $a_{\rm sym}$ by using two Wigner energy forms for all
even-even nuclei with $A$=80. The obvious discrepancy can be
observed by using two forms for Wigner energy. The solid triangles
and solid circles denote the results with form (1) and form (2),
respectively. The value of the extracted symmetry-energy
coefficients $a_{\rm sym}$ is larger with form (2) than that with
form (1), especially for the range of $I$ close to zero, and the
discrepancy decreases with increasing isospin asymmetry $I$. It is
therefore necessary to determine the Wigner energy of nuclei for a
better description of symmetry energy coefficient.

\begin{figure}
\includegraphics[angle=-0,width= 0.5\textwidth]{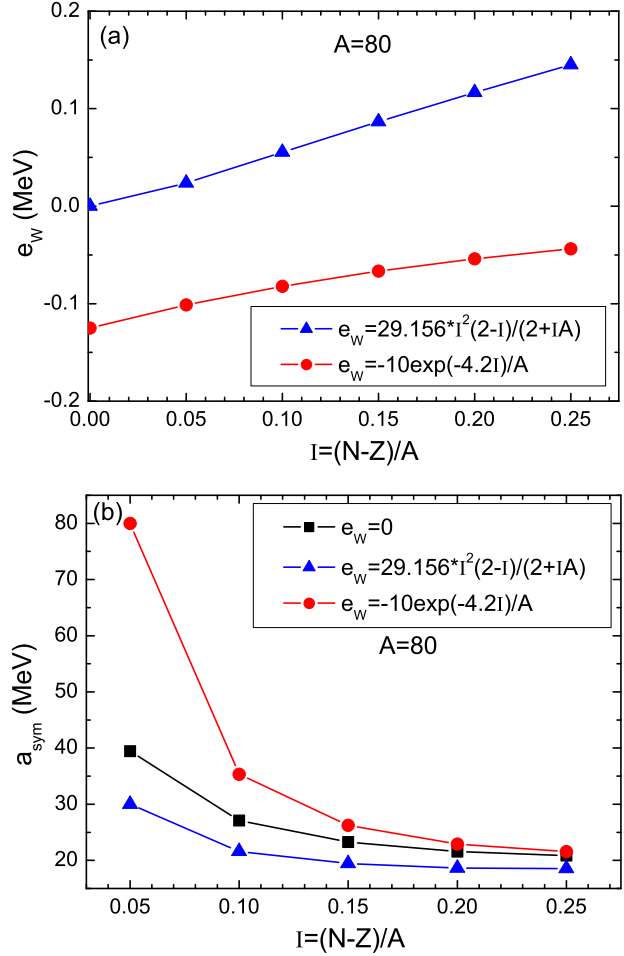}
 \caption{(Color online) (a) Two forms Wigner energy as a function of isospin asymmetry $I$, and (b) the extracted symmetry-energy coefficients $a_{\rm sym}$ by using two forms Wigner energy applied to all even-even nuclei with $A$=80. The solid squares denote the result of excluded Wigner energy.}
\end{figure}

\section{\label{sec:level1}Theoretical framework}
In semi-empirical mass formulas the Wigner energy usually is decomposed into two
parts \cite{Moller92,Satula97}
\begin{eqnarray}
E_{W}(N,Z)=-W(A)|N-Z|-d(A)\delta_{N,Z}\pi_{np},
\end{eqnarray}
where $W(A)$ and $d(A)$ are smooth functions of the nuclear mass number A. The first
term on the right-hand side of Eq. (8) contributes to all $N\neq Z$ nuclei. The quantity $\pi_{np}$ equals 1
for odd-odd nuclei and vanishes otherwise, and therefore the second term $d(A)$ is
nonzero only for $N = Z$ odd-odd nuclei. The Wigner effect mainly stems from the first term in Eq. (8). By combining
the first term in Eq. (8), the traditional symmetry energy term $(N-Z)^2/A$ is replaced by $T(T + x)$ term \cite{Neerg09,Bent13,Bent14,Isack07}. So the odd-odd symmetric nuclei are not considered in the following calculation.
$T=|T_{z}|=\frac{|N-Z|}{2}$ is the isospin value of the nuclear ground state, and $I=(N-Z)/A$ is the isospin asymmetry of a nucleus. Then one has the relation,
\begin{eqnarray}
T=\frac{|I|A}{2}.
\end{eqnarray}
The symmetry energy term including the Wigner energy can be expressed as
\begin{eqnarray}
E_{\rm sym}^{*}(A,T) = \frac{4a_{\rm sym}}{A}T(T+x)=\frac{4a_{\rm
sym}}{A}T^{2}+\frac{4a_{\rm sym}}{A}Tx.
\end{eqnarray}
Inserting Eq. (9) into Eq.(10), we can obtain the symmetry energy per particle
expression as the function of mass number $A$ and isospin asymmetry
$I$,
\begin{eqnarray}
e_{\rm sym}^{*}(A,I)=\frac{E_{\rm sym}^{*}(A,I)}{A} = a_{\rm
sym}^{*}I^{2}=a_{\rm sym}I^{2}+\frac{2a_{\rm sym}x|I|}{A},
\end{eqnarray}
where $e_{\rm sym}^{*}(A,I)=e_{\rm sym}(A,I)+e_{W}$ and $a_{\rm
sym}^{*}=a_{\rm sym}(1+\frac{2x}{|I|A})$.
$a_{\rm sym}$ is the symmetry energy
coefficient be expressed as a function of mass number $A$ and
isospin asymmetry $I$. $2a_{\rm sym}x$ denotes the Wigner energy
coefficient, the value of $x$ is not well determined from nuclear
masses, $x=1$ is associated with neutron-proton exchange
interactions in SU(2) symmetry, while $x=4$ corresponds to the full
supermultiplet symmetry SU(4)\cite{wigner37}. The further discussion
on the Wigner energy can be found in Ref. \cite{Zhao15,Fraue99,Neerg02,Neerg03}. Here $x$ as
a parameter is introduced, named the Wigner energy parameter. The $x$ value has crucial effect on the symmetry energy coefficient, since the symmetry energy
is the summation of the traditional symmetry energy and the Wigner
energy. The different $x$ value denotes the different Wigner energy.
Inserting Eq. (11) into Eq. (3) and $e_{\rm sym}^{*}(A,I)$ replacing $e_{\rm sym}(A,I)$,
the nuclear energy part per particle Eq.(3) becomes
\begin{eqnarray}
\label{eq12}
e_{n}(A,I)&=&e_{0}(A)+e_{\rm sym}^{*}(A,I)\nonumber\\[1mm]
&=&e_{0}(A)+a_{\rm sym}(A,I)(1+\frac{2x}{|I|A})I^{2},
\end{eqnarray}
Inserting Eq.(7) into Eq. (12), and take the difference of $e_{n}(A,I)$ between two isobaric nuclei with same odd-even parity. Eq. (4) becomes
\begin{eqnarray}
\label{eq13}
\Delta e_{\rm sym}^{*(i)}&=&e_{n}(A,I)-e_{n}(A,I_{i})\nonumber\\[1mm]
&=&a_{\rm sym}^{(2)}(I^{2}-I_{i}^{2})+a_{\rm
sym}^{(4)}(I^{4}-I_{i}^{4})+\frac{2a_{\rm
sym}^{(2)}x}{A}(|I|-|I_{i}|)
+\frac{2a_{\rm
sym}^{(4)}x}{A}(|I|^{3}-|I_{i}|^{3}).
\end{eqnarray}
where $i$=1, 2, 3, ..., n, $a_{\rm sym}^{(2)}=J-a_{\rm ss} A^{-1/3}$. The dependence of reference nuclei ($A,I_{1}$), ($A,I_{2}$), ... , and ($A,I_{n}$) can be canceled through the summation, and the average value $\overline{\Delta e_{\rm
sym}^{*}}$ of the difference of symmetry energy can be expressed as
\begin{eqnarray}
\nonumber\overline{\Delta e_{\rm sym}^{*}}&&=\frac{1}{n}(\Delta e_{\rm sym}^{*(1)}+\Delta e_{\rm sym}^{*(2)}+...+\Delta e_{\rm sym}^{*(n)})\\
\nonumber &&=e_{n}(A,I)-\frac{1}{n}\sum_{i=1}^{n}e_{n}(A,I_{i})\\
\nonumber &&=a_{\rm sym}^{(2)}(I^{2}-\frac{1}{n}\sum_{i=1}^{n}I_{i}^{2})+a_{\rm sym}^{(4)}(I^{4}-\frac{1}{n}\sum_{i=1}^{n}I_{i}^{4})\\
&&+\frac{2a_{\rm
sym}^{(2)}x}{A}(|I|-\frac{1}{n}\sum_{i=1}^{n}|I_{i}|)+\frac{2a_{\rm
sym}^{(4)}x}{A}(|I|^{3}-\frac{1}{n}\sum_{i=1}^{n}|I_{i}|^{3}),
\end{eqnarray}
when neglecting the microscopic shell corrections of nuclei, the result
of Eq. (14) $\overline{\Delta e_{\rm
sym}^{*}}=e_{n}(A,I)-\frac{1}{n}\sum_{i=1}^{n}e_{n}(A,I_{i})$ is
obtained by the measured binding energy per nucleon of each series
isobaric nuclei compiled in AME2012. By using the expression of the right-hand side in Eq. (14) and
fitting $\overline{\Delta e_{\rm sym}^{*}}$ from more than 2200
measured nuclear binding energies, we obtain the optimal values
$J=30.25\pm0.10$ MeV, $a_{\rm ss}=56.18\pm1.25$
MeV, $a_{\rm sym}^{(4)}=8.33\pm1.21$ MeV and $x=2.38\pm0.12$
with an rms deviation of 23.42 keV.

\begin{figure}
\includegraphics[angle=-0,width= 0.95\textwidth]{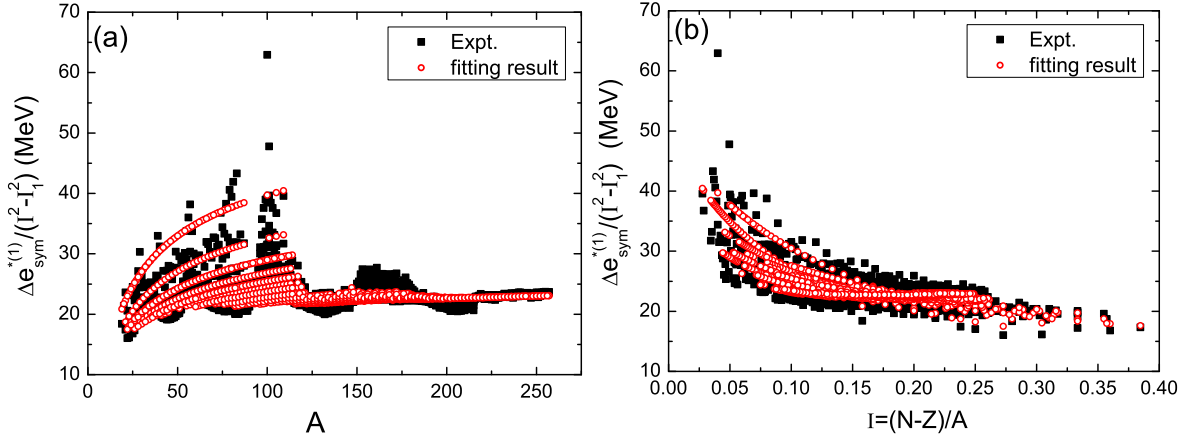}
 \caption{(Color online)Symmetry energy coefficients of nuclei as a function of (a) nuclear mass number $A$ and (b) of isospin asymmetry $I$. The solid squares and open circles denote the experimental data $\frac{\Delta e_{\rm sym}^{*(1)}}{I^{2}-I_{1}^{2}}$ and the fitting results by Eq. (14) with the optimum parameters values $J=30.25$ MeV, $a_{\rm ss}=56.18$ MeV, $a_{\rm sym}^{(4)}=8.33$ MeV and $x=2.38$.}
\end{figure}
\section{\label{sec:level1}Results and discussions}

In Fig. 3 (a), we show the extracted symmetry energy coefficients of
nuclei as a function of nuclear mass number. The solid squares
denote the extracted symmetry energy coefficients from the measured
nuclear masses by using $\frac{\Delta e_{\rm
sym}^{*(1)}}{I^{2}-I_{1}^{2}}$ in Eq. (13). The open circles denote
the fitting results by Eq. (14) with the optimum parameters values. One can see that the experimental value of $\frac{\Delta
e_{\rm sym}^{*(1)}}{I^{2}-I_{1}^{2}}$ obtained in our approach by
Eq. (13) shows some oscillations and fluctuations, which is probably
caused by the shell effects and other nuclear structure effects. In Fig. 3 (b), we show the same data as in Fig. 3 (a), but as a function of
isospin asymmetry $I$. Form Fig .1 and Fig. 3 (b), we can find that
the extracted symmetry energy coefficients depend on the
corresponding isospin asymmetry of nuclei, which decreases with
increasing isospin asymmetry $I$ for the same mass number $A$, the
largest values located in the range of nearly symmetric nuclei.
However the Wigner energy parameter $x$ value influences
every parameters in Eq. (14). Fig. 4 shows the coefficients $J$,
$a_{\rm ss}$, $a_{\rm sym}^{(4)}$ (in MeV) and $\sigma$ deviation (in
keV) as a function of Wigner energy parameter $x$. From Fig. 4 we
can see that the coefficients $J$ (solid squares), $a_{\rm ss}$ (solid
circles) and $a_{\rm sym}^{(4)}$ (solid triangles) increase firstly
then decrease with increasing $x$ values in the range from 0 to 12. The rms
deviation $\sigma$ (down triangles) decreases firstly then increases
with increasing $x$ values. The minimum value of $\sigma=23.42$ keV
is corresponding to the set optimal parameters values. One
may thus expect the coefficient $x$ to lie somewhere between 1 and
4. The volume symmetry coefficient $J\simeq 30$ MeV is insensitive
to the value $x$ in the range $1\leq x\leq 4$. The surface symmetry
coefficient $a_{\rm ss}$ is sensitive to the value $x$ in the range $1\leq
x\leq 4$, whose value changes from 38.72 MeV to 65.85 MeV. The
coefficient $a_{\rm sym}^{(4)}$ is more sensitive dependence of the
value $x$ in the range $1\leq x\leq 4$, from -6.98 MeV to 16.56 MeV. So we draw
a conclusion from the figure that $a_{\rm sym}^{(4)}$ is not well
determined from nuclear masses since $x$ is ill-determined. For
example, we change $x$ value somewhat from 1.5 to 1.6, the value of
$a_{\rm sym}^{(4)}$ changes from negative to positive. So the sign
(positive or negative) of $a_{\rm sym}^{(4)}$ is sensitively
dependent on the value of $x$.

\begin{figure}
\includegraphics[angle=-0,width= 0.85\textwidth]{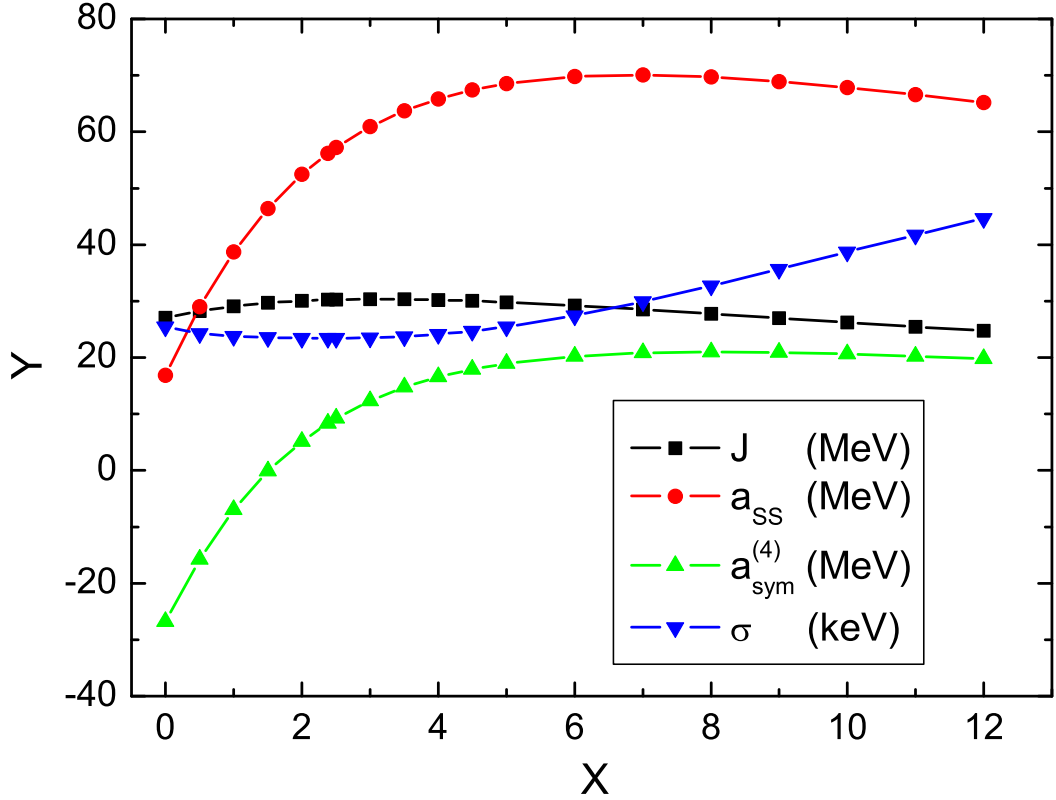}
 \caption{(Color online) The volume symmetry coefficient $J$, surface symmetry coefficient $a_{\rm ss}$, the coefficient $a_{\rm sym}^{(4)}$ of $I^{4}$ term  (in MeV) and $\sigma$ deviation (in keV) as a function of Wigner energy parameter $x$.}
\end{figure}

\begin{figure}
\includegraphics[angle=-0,width= 0.45\textwidth]{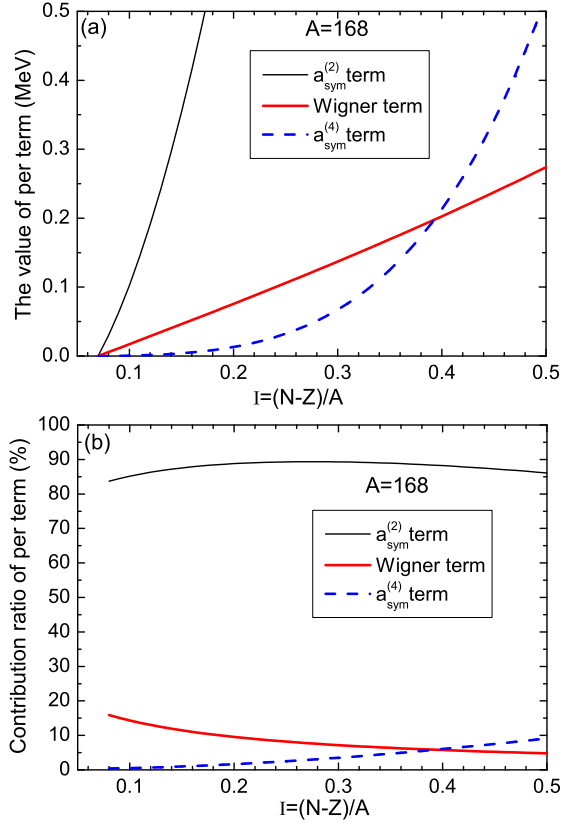}
 \caption{(Color online) (a) The values of $a_{\rm sym}^{(2)}$ term (thin curve), Wigner term (thick curve) and $a_{\rm sym}^{(4)}$ term (dashed curve) in Eq. (13) as a function of $I$, and (b) the contribution ratio of per term for $A=168$. The parameters of $J=30.25$ MeV, $a_{\rm ss}=56.18$ MeV, $a_{\rm sym}^{(4)}=8.33$ MeV and $x=2.38$ are used.}
\end{figure}

The contributions of symmetry energy and Wigner energy are also studied. As an example, the contribution of per term is shown in Fig. 5, where the asymmetric nucleus $I_{1}=0.07$ is
selected as reference nucleus, since it is the minimum value of known nuclei in mass table
AME2012 for $A=168$. From Fig. 5(a) one can see that the value of
all three term increase with increasing isospin asymmetry $I$, when
$I<0.39$, the value of $a_{\rm sym}^{(4)}$ term $a_{\rm sym}^{(4)}(I^{4}-I_{1}^{4})$ is less
than that of Wigner term $\frac{2x}{A}[a_{\rm sym}^{(2)}(|I|-|I_{1}|)+a_{\rm sym}^{(4)}(|I|^{3}-|I_{1}|^{3})]$ in Eq. (13), when
$I\geq0.39$ the value of $a_{\rm sym}^{(4)}$ term  is
larger than that of Wigner term. Fig. 5(b) shows the contribution ratio of per term, the ratio is
calculated by the ratio of per term value to the summation of three
term value. From Fig. 5(b) we can see the changing details of per
term with increasing isospin asymmetry $I$. The average contribution
ratio of four term are 87.92\%, 8.27\% and 3.81\% for $a_{\rm sym}^{(2)}$ term
$a_{\rm sym}^{(2)}(I^{2}-I_{1}^{2})$, Wigner term and $a_{\rm sym}^{(4)}$ term in the range of
$I=0.07-0.5$, respectively. With the increasing of isospin asymmetry
$I$, the $a_{\rm sym}^{(2)}$ term is the major contributor, which increases firstly
and reaches a maximum at $I=0.27$, and then decreases with
increasing isospin asymmetry $I$. The Wigner term decreases and the
$a_{\rm sym}^{(4)}$ term increases with increasing isospin asymmetry $I$. The
contribution ratio of $a_{\rm sym}^{(4)}$ term is less
than that of Wigner term in the range
of $I=0.07-0.39$ and larger than that when $I\geq0.39$.

\section{\label{sec:level1}Summary}

In summary, we have proposed a method to extract the symmetry energy
coefficient (including the coefficient $a_{\rm sym}^{(4)}$ of
$I^{4}$ term) from the differences of available experimental binding
energies of isobaric nuclei. The advantage of this approach is that
one can efficiently remove the volume, surface and pairing energies
in the process. It is found that the extracting experimental symmetry energy
$e^{*}_{\rm sym}(A,I)$ should be the summation of the traditional
symmetry energy $e_{\rm sym}(A,I)$ and the Wigner energy
$e_{W}(A,I)$. And $a^{*}_{\rm sym}(A,I)$ decreases with increasing of isospin
asymmetry $I$, which is mainly caused by the Wigner energy effect.
Through the polynomial fit to the result of $\overline{\Delta e_{\rm
sym}^{*}}$ by the right-hand side expression of Eq. (14), we have
obtained the optimum parameters values $J=30.25\pm0.10$ MeV, $a_{\rm ss}=56.18\pm1.25$
MeV, $a_{\rm sym}^{(4)}=8.33\pm1.21$ MeV and the Wigner parameter $x=2.38\pm0.12$.
We also find that the volume symmetry coefficient
$J\simeq 30$ MeV is insensitive to the value $x$, while the
surface symmetry coefficient $a_{\rm ss}$ and the coefficient $a_{\rm
sym}^{(4)}$ are very sensitive dependence of the value $x$ in the
range $1\leq x\leq 4$, especially for $a_{\rm sym}^{(4)}$, whose
value maybe change from negative to positive since the change $x$
value somewhat in the range 1 to 4. The contribution
of the wigner energy term decreases and the contribution of $a_{\rm
sym}^{(4)}$ term increases with increasing of isospin asymmetry $I$.
For very neutron-rich nuclei, $a_{\rm
sym}^{(4)}$ term will play an important role since its contribution is larger than that of Wigner energy term.

\begin{center}
\textbf{ACKNOWLEDGEMENTS}
\end{center}
We thank Dr. H. Jiang for helpful communications. This work was
supported by National Natural Science Foundation of China, Nos.
11475004, 11275052, 11305003, 11375094 and 11465005, the Natural Science
Foundation of He'nan Educational Committee Nos.2011A140001 and
2011GGJS-147, and innovation fund of undergraduate at Anyang Normal
University (ASCX/2014-Z57). N. W. acknowledges the support of the
Open Project Program of State Key Laboratory of Theoretical Physics,
Institute of Theoretical Physics, Chinese Academy of Sciences, China
(No. Y4KF041CJ1).

\end{document}